\documentclass[preprint,showpacs,onecolumn,nofootinbib]{revtex4}
\usepackage[colorlinks=true,linkcolor=blue,urlcolor=blue,filecolor=black,citecolor=red,pdfstartview=FitV,pdftitle={},pdfsubject={},pdfkeywords={},bookmarksopen=true]{hyperref}
\usepackage{graphicx}
\usepackage{amsmath}
\usepackage{amsfonts}
\usepackage{bookmark}
\usepackage{amssymb,ulem}
\usepackage{color}%
\usepackage{dcolumn}

\usepackage{MnSymbol,wasysym}
\usepackage{braket}
\usepackage{color}%
\usepackage{eurosym}
\usepackage{calrsfs}
\usepackage[usenames,dvipsnames,svgnames]{xcolor}

\newcommand{\RNum}[1]{\uppercase\expandafter{\romannumeral #1\relax}}

\begin{document}
\baselineskip=0.6 cm
\title{Holographic entanglement entropy and complexity in St$\ddot{u}$ckelberg superconductor}
\author{Hong Guo$^{1,2}$}
\email{gh710105@gmail.com}
\author{Xiao-Mei Kuang$^{3,4}$}
\email{xmeikuang@yzu.edu.cn}
\author{Bin Wang$^{3,4}$}
\email{wang$_$b@sjtu.edu.cn}
\affiliation{\small{$^1$~Department of Physics and Astronomy, Shanghai Jiao Tong University, Shanghai 200240, China}}
\affiliation{\small{$^2$~Collaborative Innovation Center of IFSA (CICIFSA), Shanghai Jiao Tong University, Shanghai 200240, China}}
\affiliation{\small{$^3$~Center for Gravitation and Cosmology, College of Physical Science and Technology, Yangzhou University, Yangzhou 225009, China}}
\affiliation{\small{$^4$~School of Aeronautics and Astronautics, Shanghai Jiao Tong University, Shanghai 200240, China}}

\date{\today }

\begin{abstract}
\baselineskip=0.6 cm
The holographic superconductors, as one of the most important application of gauge/gravity duality, promote the study of
strongly coupled superconductors via classical general relativity living in one higher dimension. One of the interesting
properties in holographic superconductor is the appearance of first and second order phase transitions. Recently, another
active studies in holographic framework is the holographic entanglement entropy and complexity evaluated from gravity side. In this note,
we study the properties of the holographic entanglement entropy and complexity crossing both first and second order phase transitions in St$\ddot{u}$ckelberg superconductor.
We find that they behave differently in two types of phase transitions. We argue that  holographic entanglement entropy and complexity conjectured with the volume
can also be a possible probe to the type of superconducting phase transition.
\end{abstract}

\maketitle

\section{Introduction}
In the last decades, the AdS/CFT correspondence \cite{Maldacena:1997re,Gubser:1998bc,Witten:1998qj}, has made significant progress and development. In many applications of gauge/gravity duality, AdS and condensed matter physics duality (AdS/CMT) has attracted plenty of interest, which provides a powerful approach to study the strongly coupled field theory in terms of weakly coupled gravitational systems. In particular, physicist has investigated kinds of  holographic superconductors which was first proposed in \cite{Gubser:2008px,Hartnoll:2008vx} including s-wave model \cite{Hartnoll:2008kx}, p-wave model \cite{Gubser:2008wv} and d-wave model \cite{Chen:2010mk,Benini:2010pr}.
Considerable efforts  in holographic superconductors have been made in the related research and discussion, and for reviews, readers can see for example, \cite{Horowitz:2010gk,Cai:2015cya} and references therein.

In the study of holographic superconductors and other condensed matter physical models like holographic superfluidity,  many study on the types of phase transition analysis have been done. The first proposal on holographic s-wave superconductor in probe limit  was found to undergo a second order phase transition from normal state\cite{Hartnoll:2008vx}. Later,
It was addressed  in \cite{Franco:2009if}   that  the holographic superconductor via the St$\ddot{u}$ckelberg mechanism allows
the first order phase transition to occur when the model parameter surpasses a critical  value, which was extended in \cite{Franco:2009yz,Ma:2011zze,Hafshejani:2018svs}. First order phase transition in holographic s-wave superconductor has also been observed in  superfluidity  model\cite{Herzog:2008he}, $N = 8$ gauged supergravity model\cite{Albash:2012pd} and so on.
Besides, in p-wave holographic superconductor models, some works for instance \cite{Cai:2013pda,Cai:2013aca,Li:2013rhw,Giordano:2015vsa} showed  the zero-order, first-order and second-order phase transition in their models, and discovered the retrograde condensation in certain parameter space. 

On the other hand, holographic entanglement entropy\cite{Ryu:2006bv} as a measure  of the degrees of freedom in
a strongly coupled system, has been firstly evaluated in holographic metal/superconductor phase transition in \cite{Albash:2012pd}.
It was addressed that  the entanglement entropy in superconducting case is always less than that in the normal phase. Beside the entanglement entropy is continuous but its slope in terms of
temperature is discontinuous  at the transition temperature $T_c$ for the second order phase transition,  while for the first order phase transition, the entanglement entropy presents
a discontinuous drop at the critical temperature. Thus, the authors of \cite{Albash:2012pd} argued  that the holographic entanglement
entropy can be used to determine the orders of superconducting phase transition. Later, the authors of \cite{Kuang:2014kha}  found that the entanglement entropy has a different behavior near the contact
interface of the superconducting to normal phase  due to the proximity effect. Further effort in using holographic entanglement entropy as a probe of phase transition has been made in
\cite{Cai:2012nm,Das:2017gjy,Peng:2014ira,Yao:2014fwa,Yao:2016ils,Dey:2014voa,Romero-Bermudez:2015bma,Peng:2015yaa,Momeni:2015iea,Ling:2016wyr,Ali:2018aon} and therein. The authors studied  the behaviors of holographic entanglement entropy in different orders phase transition and also in quantum phase transition. They  argued that the  holographic entanglement entropy  may not be universal in holographic superconduction models, which  explains the existence of the difference between in first order phase transition and second order phase transition.

In this paper, we will investigate the properties of holographic complexity measured by the volume\cite{Alishahiha:2015rta}  in St$\ddot{u}$ckelberg holographic superconductor model. The  complexity essentially measures the difficulty of turning a quantum state into another and so  it could reflect a phase transition on the boundary field theory.  Holographic complexity has been studied in one dimensional s-wave superconductor in  \cite{Momeni:2016ekm,Zangeneh:2017tub,Chakraborty:2019vld}, p-wave superconductor in \cite{Fujita:2018xkl} and QCD phase transition\cite{Ghodrati:2018hss}. It was found that holographic complexity behave in the different way with holographic entanglement entropy and both of them can reflect the one dimensional phase transitions.

Here we are aiming to study the properties of  holographic complexity for a strip subregion when the system crosses the first order and second order  St$\ddot{u}$ckelberg superconducting phase transition. We will consider only the $\psi^4$ coupling term in  St$\ddot{u}$ckelberg model, which brings in first order phase transition when the coupling parameter is bigger than a critical value in probe limit case \cite{Franco:2009if} and in backreaction case\cite{Peng:2014ira}. We will see that the holographic  complexity increases monotonously  as the temperature  becomes lower for  the second order phase transition, while it present a multivalue area in high temperature region in first order case which  appears in the same temperature region in free energy diagram. Besides, similar to the holographic entanglement entropy, at the second order phase transition temperature, the complexity is continuous and its slope in terms of the temperature have a jump, while in the first order phase transition case, the  complexity presents a jump at the critical temperature. Thus, we are expecting that similar to the holographic entanglement entropy, the holographic complexity can also be a good probe to the order of superconducting phase transition.

Our paper is organized  as follows. We briefly show the St$\ddot{u}$ckelberg superconducting phase transition, the  condensation and  computing the free energy of the dual system in section \ref{sec-supercon}. In section \ref{sec-HEEhc}, we first review the holographic setup of entanglement  entropy and complexity for a strip subregion in AdS/CFT framework, and then numerically evaluate them in the superconducting model. The last section contributes to our conclusion and discussion.

\section{St$\ddot{U}$ckelberg Superconducting phase with backreaction}\label{sec-supercon}
The generalized four dimensional action containing a $U(1)$ gauge field and the scalar field coupled via a
generalized St$\ddot{u}$ckelberg Lagrangian is \cite{Franco:2009yz}
\begin{eqnarray}\label{lagrange-1}
S&=&\int
d^4x\sqrt{-g}\left[\left(R+\frac{6}{l^{2}}\right)+L_{M}\right],
\end{eqnarray}
where $l$ is the AdS radius which will be set to be unity in the
following discussion. $L_{M}$ is the generalized
St$\ddot{u}$ckelberg Lagrangian
\begin{eqnarray}\label{lagrange-2}
L_{M}&=-\frac{1}{4}F^{\mu\nu}F_{\mu\nu}-(\partial{\psi})^{2}
 -m^{2}|{{\psi}}|^{2} -|\mathcal{F}(\psi)|(\partial{p}-q A)^{2},
\end{eqnarray}
where $\mathcal{F}(\psi)$ is a function of $\psi$. In this paper, we will consider $\mathcal{F}(\psi)=\psi^{2}+c_{4}\psi^{4}$ where $c_{4}$ is a model parameter\footnote{We note that the holographic superconductor  with $\mathcal{F}(\psi)=\psi^{2}+c_{6}\psi^{4}$ coupling with backreaction has been studied in \cite{Cai:2012es}.}. Considering the gauge symmetry $\tilde{A}_{\mu}\rightarrow~\tilde{A}_{\mu}+\partial\Lambda$~and~$p\rightarrow~p+\Lambda$,
we fix the gauge $p = 0$ by using the gauge freedom.

To include the backreaction, we consider the metric ansatz
\begin{eqnarray}\label{metric}
ds^{2}&=&\frac{-f(z)}{z^2}e^{-\chi(z)}dt^{2}+\frac{dz^{2}}{z^2 f(z)}+\frac{dx^{2}+dy^{2}}{z^{2}}.
\end{eqnarray}
The Hawking temperature is expressed as
\begin{equation}
T=-\frac{f'(z_{H})e^{-\chi(z_{H})/2}}{4\pi},
\end{equation}
and the event horizon $z_{H}$ satisfies $f(z_{H})=0$.
Then, considering the ansatz of  the matter fields as $\psi=\psi(z)~,~A=\phi(z) dt$, we obtain the
equations of motion from the action \eqref{lagrange-1} under the metric \eqref{metric}
\begin{eqnarray}\label{BHeom-1}
\psi''-\left(\frac{2}{z}+\frac{\chi'}{2}-\frac{f'}{f}\right)\psi'+\frac{q^{2}\phi^{2}e^{\chi}}
{f^{2}}\left(\psi+2q^{2}c_{4}\psi^{3}\right)-\frac{m^{2}}{z^2 f}\psi=0,&&\\
\label{BHeom-2}
\phi''+\frac{\chi'}{2}\phi'-\frac{q^{2}\phi}{z^2 f}\left(\psi^{2}+q^{2}c_{4}\psi^{4}\right)=0,&&\\
\label{BHeom-3}
\chi'-\frac{z}{2}\left(\psi'^{2}+\frac{q^{2}\phi^{2}e^{\chi}}{f^{2}}\left(\psi^{2}+q^{2}c_{4}\psi^{4}\right)\right)=0,&&\\
\label{BHeom-4}
f'-\frac{zf}{4}\left(\psi'^2+\frac{12}{z^2}+\frac{z^2 e^{\chi}}{f}\phi'^{2} \right)-\frac{\psi^2}{4z}\left(m^2+\frac{z^{2}q^{2}\phi^{2}e^{\chi}}{f}\left(\psi^{2}+q^{2}c_{4}\psi^{4}\right)\right)+\frac{3}{z}=0.
\end{eqnarray}

In our following study, we will take $q=1$ and $m^2=-2$ without loss of generality,  though our analysis can be extended into other proper parameters. Thus, the behavior of various field in the asymptotical AdS boundary is
\begin{eqnarray}\label{infboundary}
&&\psi=\psi_{1}z+\psi_{2}z^2\cdot\cdot\cdot,\
\phi=\mu-\rho z+\cdot\cdot\cdot,
f=1+f_3 z^3\cdot\cdot\cdot,
\chi=\chi_0+\chi_1z\cdot\cdot\cdot.
\end{eqnarray}
According to AdS/CFT dictionary, $\mu$ and $\rho$ are considered as the chemical potential and charge of the boundary field theory, respectively.  $\psi_{1}$ and $\psi_{2}$ can be dual to the source while the other is the vacuum expectation value due to the choice of standard quantization or alternative one.  As is pointed out in \cite{Hartnoll:2008kx} that $\psi_{1}$ as expectation value is always divergent at zero temperature limit, i.e., it may be not physical. So we will choose $\psi_{1}$ as the source and $\psi_{2}$ is related to the expectation value $\langle O_2\rangle~$. Moreover, near the horizon, the regular condition requires that the Maxwell field satisfies $\phi(r_{H})=0$ and the scalar field is not vanished.

\begin{figure}[h]
\center{
\includegraphics[scale=0.6]{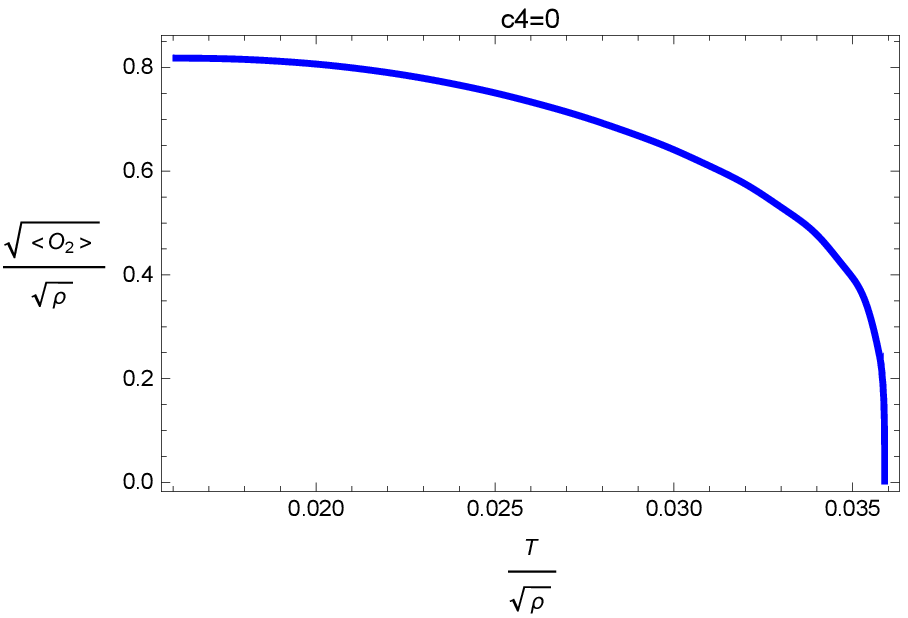}\hspace{1cm}
\includegraphics[scale=0.6]{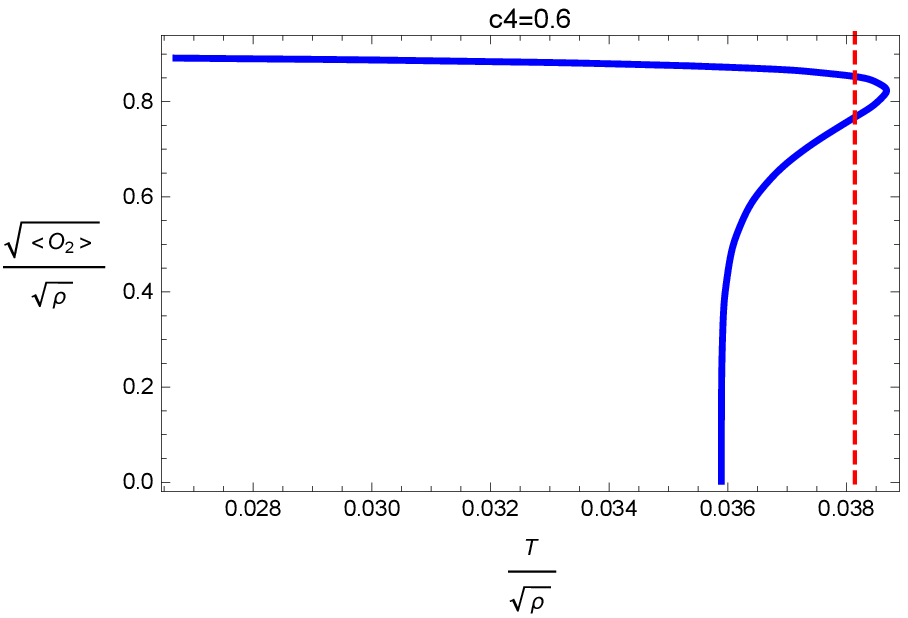}
\caption{\label{figcond} The condensates of the
scalar operator $\langle O_2\rangle$ for $c_4=0$ and $c_4=0.6$. The left curve describes the second-order phase transition and the right multi-value curve describes the first-order phase transition.}}
\end{figure}

We numerically solve the equations of motion \eqref{BHeom-1}-\eqref{BHeom-4} by setting the source $\psi_{1}=0$.  Our numerical results  of the condensation are explicitly shown in figure \ref{figcond} which indicates that  the type of phase transition is affected by the values of coupling parameter $c_4$.  There exists a critical value of $c_{4c}\sim 0.5$ below which the system undergoes a second superconducting phase transition while above which a first order phase transition  occurs in the system. Similar phenomena  was also found in \cite{Franco:2009if}.

For the second order phase transition, the critical temperature $T_c$ is straightforward to be read off, at which the condensation appears. We find that the critical temperature is around  $0.0358912$ which is not explicitly dependent on coupling parameter. This is an obvious result indicating that the physical process is stable when the system has the second-order phase transition. Similar phenomena have also been found in the studies\cite{Franco:2009if,Peng:2014ira}.
However, for the first order phase transition, it is not that direct to fix $T_c$ because of the multivalue of the condensation. Then, a usual way to fix it is to calculate the free energy of the solutions.

In order to further determine  $T_c$ for the phase transitions and its order, we will compute the free energy of the system. We will work in the  canonical ensemble, in which the charge is fixed. According to \cite{Balasubramanian:1999re,Skenderis:2002wp}, the free energy of the boundary field theory is connected with on-shell Euclidean bulk action as $F=-T S_E$ and in our model
\begin{eqnarray}
S_E&=&-(S_{EH}+S_{\Psi}+S_A)+\int_{z\to 0}dx^3\sqrt{-g_{\infty}}(-2K+4/L^2),
\end{eqnarray}
where $g_{\infty}$ is the induced metric on the boundary, and $K$ is the trace of extrinsic curvature. Then considering the equations of motion \eqref{BHeom-1}-\eqref{BHeom-4} and the boundary behaviors \eqref{infboundary}, the free energy density in our model is evaluated by
\begin{equation}
\frac{F}{V_2}=f_3
\end{equation}
where $V_2=\int dxdy$ which will be set to be unit and $f_3$ is a coefficient in \eqref{infboundary}. We notice that for the normal state,  the free energy is $f_3=\mu/4-1$.

\begin{figure}[h]
\center{
\includegraphics[scale=0.6]{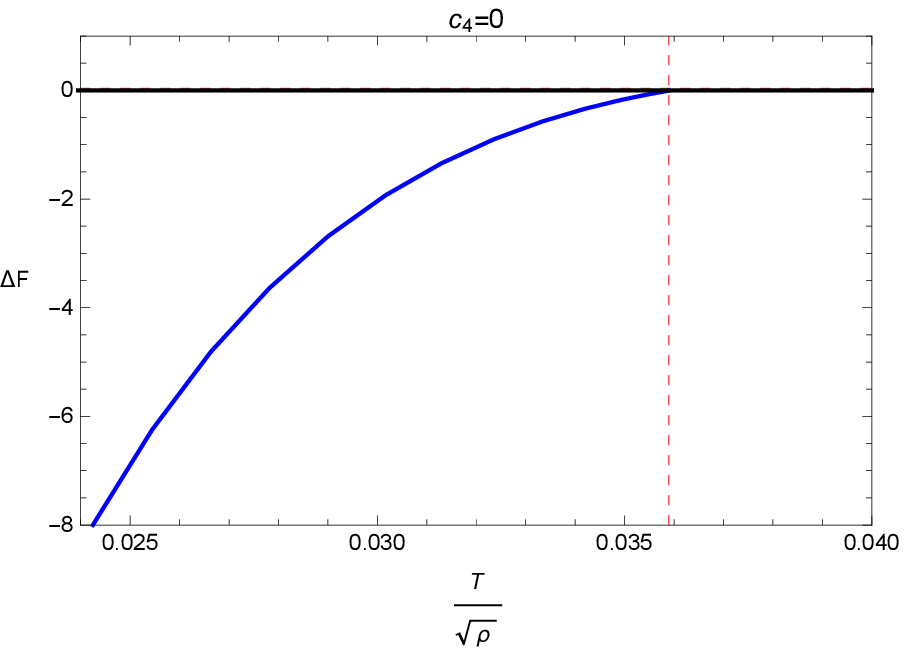}\hspace{1cm}
\includegraphics[scale=0.6]{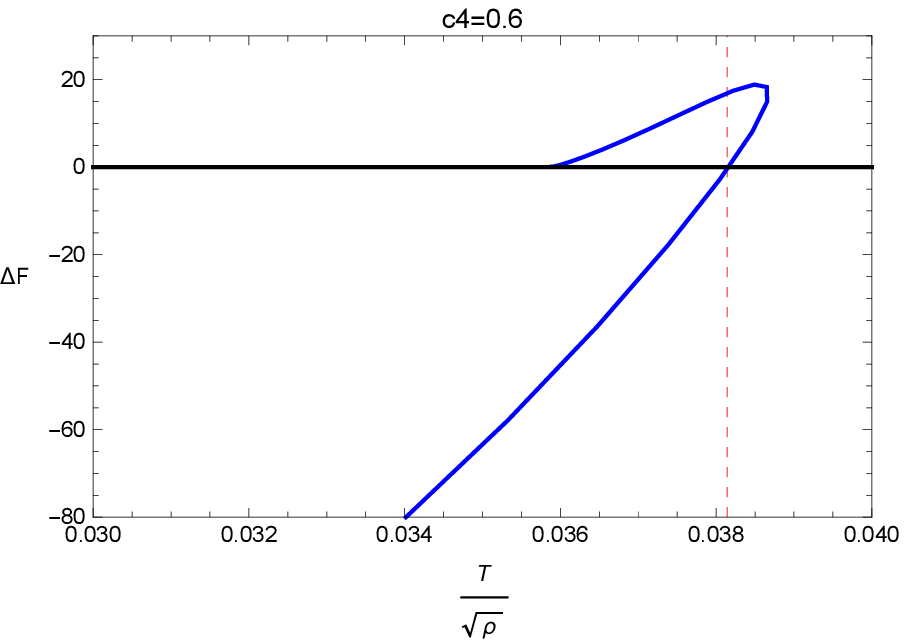}
\caption{\label{figfree} The free energy of the system for $c_4=0$ and $c_4=0.6$.}}
\end{figure}

The results of $\Delta F=F-F_{RN}$, which is the difference between the free energy of the condensed state and normal state,  are shown in figure \ref{figfree}. The blue solid lines are for the superconducting state  while the black lines represent the normal state.  For $c_4=0$ in the left plot, it is explicit that when  $T>T_c$, the RN solution is physically favorable;  when $T=T_c$  which  is marked by the red dashed vertical line, second order phase transition occurs and the superconducting phase become physically favorable as the temperature further decreases.  For $c_4=0.6$ in the right plot,  the multivalue region of the $\Delta F$ appears which implies that first order phase transition occurs. Though the temperature $T=0.03559$ represents the appearance of condensation function, but  physical phase transition happens at the interaction point $T\simeq 0.038143$, below which the superconducting state is thermodynamically favorable.

Thus, from the  free energy, we can  read off $T_c$ for the first order transition for different $c_4$ which are listed  in table \ref{table-Tc1}.  $T_c$  for the first order phase transition grows as  the coupling parament, which behaves differently with that in the  second order phase transition. Comparing the two types of the superconducting phase transition, the second-order phase transitions has lower but stable critical temperatures, so more drastic condensation and phase transition process is hinted in the first order case.

\begin{table}[h]
\center{
\begin{tabular}{|c|c|c|c|c|c|}\hline
$c_4$&$0.52$&$0.55$&$0.60$&$0.70$&$0.80$
\\\hline
$T_c$&$0.0365796$&$0.0372203$&$0.0381431$&$0.0403314$&$0.0423508$
\\\hline
 \end{tabular}
\caption{\label{table-Tc1} The critical temperature for the first order phase transition which grows with the coupling parameter.}}
\end{table}

Next, we shall study holographic entanglement entropy and holographic complexity in the two types of phase transitions. We expect that the holographic entropy and complexity would perform different properties in the two types of phase transition, and suggest more deeply physics about the superconducting phase transitions in the dual systems.

\section{Holographic entanglement entropy and holographic complexity}\label{sec-HEEhc}
\subsection{The setup}
\begin{figure}
\center{
\includegraphics[scale=0.2]{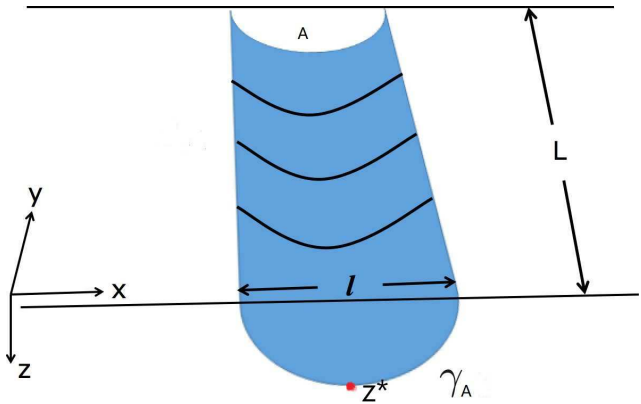}\hspace{0cm}
\caption{\label{figstripe} Cartoon of the strip geometry. $z$ is the radial coordinate in the bulk spacetime. In the duality, the boundary theory lives at $z=0$ and $z^{\star}$ is the turning point of the minimal surface $\gamma_{A}$. }}
\end{figure}

We  consider the subsystem $A$ with a straight strip geometry which is described by $-\frac{l}{2}\leq x \leq\frac{l}{2}\,, 0\leq y \leq L$(see Fig.\ref{figstripe}). Here $l$ is  the size of region $A$ and $L$ is a regulator which can be set to be infinity. Recalling the proposal by Ryu and
Takayanagi\cite{Ryu:2006bv}, the radial minimal extended surface $\gamma_A$ bounded by $A$ with a turning point $z_{\star}$ determine the entanglement entropy $S_A$ as via
 \begin{equation}\label{RTF}
S_A=\frac{\mbox{Area}(\gamma_A)}{4G^{(d+2)}_N}\ .
\end{equation}
In holographic framework, there are two different proposals on how to evaluate the complexity.
One is the CV conjecture (Complexity=Volume) \cite{Stanford:2014jda,Susskind:2014jwa}.
Another is the CA conjecture (Complexity=Action) \cite{Brown:2015bva,Brown:2015lvg}.
CV conjecture is that the holographic complexity is proportional to the volume of
a codimension-one hypersurface with the AdS boundary and the Holographic Ryu-
Takayanagi surface.
While in CV conjecture, one identifies the HC with the gravitational action evaluated on the Wheeler-DeWitt patch in the bulk. Here we will employ CV conjecture and following Alishahiha's proposal\cite{Alishahiha:2015rta}, the holographic complexity for this system $ A$
is measured by the volume enclosed by $\gamma_A$ as
 \begin{equation}\label{Comp}
C=\frac{\mbox{Volume}(\gamma_A)}{8G^{(d+2)}_N}\ ,
\end{equation}
where  $G^{(d+2)}_N$ is the Newton constant of the general gravity in AdS$_{d+2}$ theory.

Following \cite{Albash:2012pd},
the induced metric of the hypersurface $\gamma_A$ who has  the same boundary as the stripe does is
\begin{equation}
ds_{induced}^2=\frac{1}{z^2}\Big[(\frac{1}{f}+x'(z))dz^2+dy^2\Big]~.
\end{equation}
Then, the holographic entanglement entropy connecting with the area of the surface can be evaluated  as
\begin{equation}\label{entanglement1}
4G_{4}S=Area(\gamma_A)=\int dy \int_{-\ell/2}^{\ell/2} \frac{dx}{z^2}\sqrt{1+\frac{1}{f(z)}(\frac{dz}{dx})^2}.
\end{equation}
The condition $\left(\frac{dz}{dx}\right)^2=f(z)\left(\frac{z_{\ast}^4}{z^4}-1\right)$  gives the minimal area  in  (\ref{entanglement1})
where $z_{\ast}$ is the location in $z$ with the smooth extremal surface\cite{Ben-Ami:2016qex}. Integrating the condition gives us
\begin{eqnarray}\label{x1}
	x(z)=\int^{z_{\ast}}_{z} \frac{z^2dz}{\sqrt{(z_{\ast}^4-z^4)f(z)}}.
\end{eqnarray}

Subsequently, the entanglement entropy \eqref{RTF} is geometrized in terms of AdS/CFT dictionary as
\begin{equation}\label{entanglement2}
S=\frac{L}{2G_{4}}\int^{z_{\ast}}_{\epsilon} \frac{z_{\ast}^2}{z^2}\frac{dz}{\sqrt{(z_{\ast}^4-z^4)f(z)}}=\frac{L}{2G_{4}}(\frac{1}{\epsilon}+s),
\end{equation}
where $L=\int dy$ and $\epsilon\rightarrow0$ is the UV cutoff.
And according to \cite{Alishahiha:2015rta,Momeni:2016ekm}, the holographic  complexity  \eqref{Comp} is
holographically related to the volume in the bulk enclosed by
$\gamma_A$ as
\begin{eqnarray}\label{comple}
	C=\frac{1}{4\pi G_{4}}\int_{\epsilon}^{z_{\ast}}dz\frac{x(z)}{z^3\sqrt{f(z)}},
\end{eqnarray}
where $x(z)$ is expressed in \eqref{x1}.

It is noticed that from the definitions \eqref{RTF} and \eqref{Comp},  one may expect that the
analytic  features of entanglement entropy and complexity should be similar because they are both determined in terms of the analytic features of the extremal geometry. Specially, according to the formulas \eqref{entanglement2} and \eqref{comple}, it is obvious that the profile of $f(z)$ mainly determines the analytic features, which we will study in detail later.  However,
the entanglement entropy is usually not enough to describe the rich geometric structure because it grows in a very short time during the thermalization process of a strongly coupled system\cite{Susskind:2014moa}. So it is still significant to study the complexity which may help us to enhance and supplement the description of the phase transition in the holographic superconductor model. Next, we will present the behavior of the two physical quantities numerically.

\subsection{The results}
We shall  analyze the properties of the holographic entanglement entropy and complexity of the  dual system in different types of superconducting phase transitions.

\begin{figure}[h]
\center{
\includegraphics[scale=0.6]{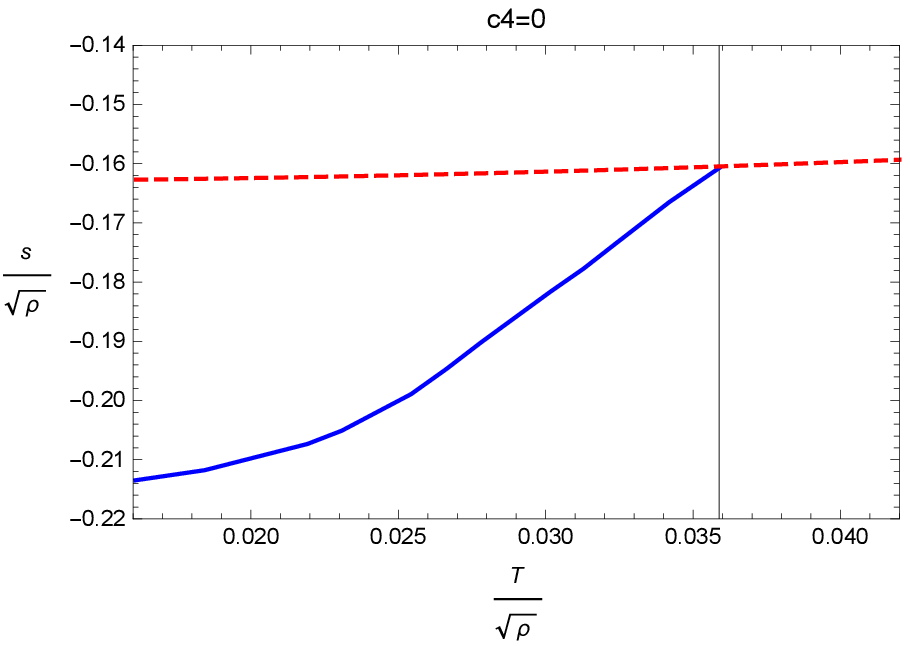}\hspace{1cm}
\includegraphics[scale=0.6]{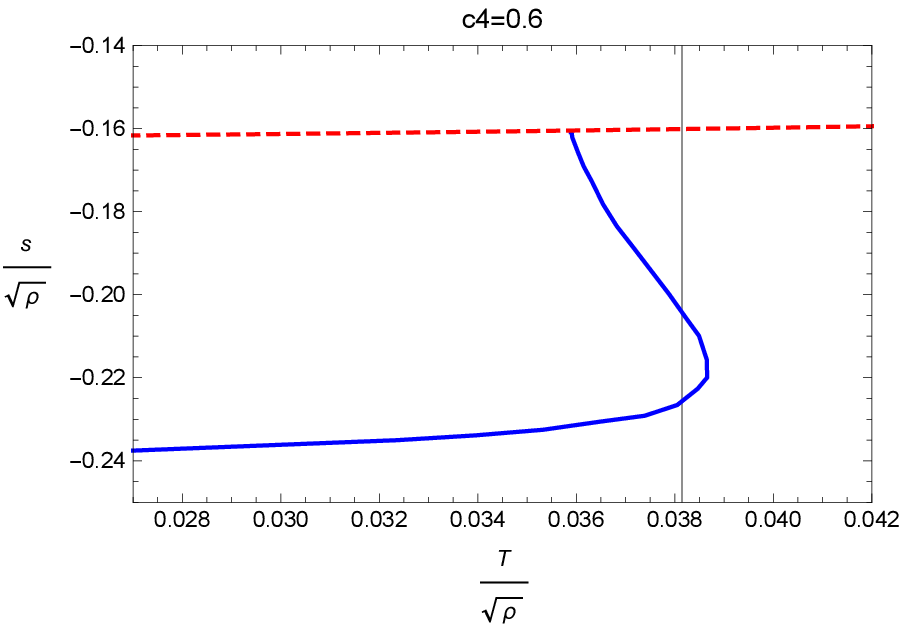}
\caption{\label{fighhe} The holographic entanglement entropy behaves as the temperature for $c_4=0$ and $c_4=0.6$.}}
\end{figure}

The numerical results of holographic entanglement entropy are shown in figure \ref{fighhe} where the red  dashed lines are for the normal state and the blue lines are for the condensed state.  For the second order phase transition with  $c_4=0$ (left plot), $s$ decreases monotonously as the temperature becomes lower. When the temperature reaches the critical temperature $T_c=0.0358912$, the holographic entropy separates into two branches which implies a phase transition.  And the values for  the superconducting state is always lower than that for  the normal state. This is reasonable because the cooper pairs form in the superconducting state which suppress of the degree of freedom of the system. Contrasting to the monotonous behavior of second order phase transition, the entanglement entropy are always non-monotonous for the first order phase transition with $c_4=0.6$(right plot), and has  a discontinuous drop to the minimal entropy  at  the critical temperature $T_c=0.038143$. Similar phenomena has also been found in the first order of insulator/superconducting phase transition\cite{Cai:2012es,Peng:2014ira}. We note that we always expect new
degrees of freedom to emerge in new phases, so that  the discontinuity or drop of the entropy at transition point  may imply  non-trivial  reorganization of the degrees of freedom in the system.

\begin{figure}[ht]
\center{
\includegraphics[scale=0.6]{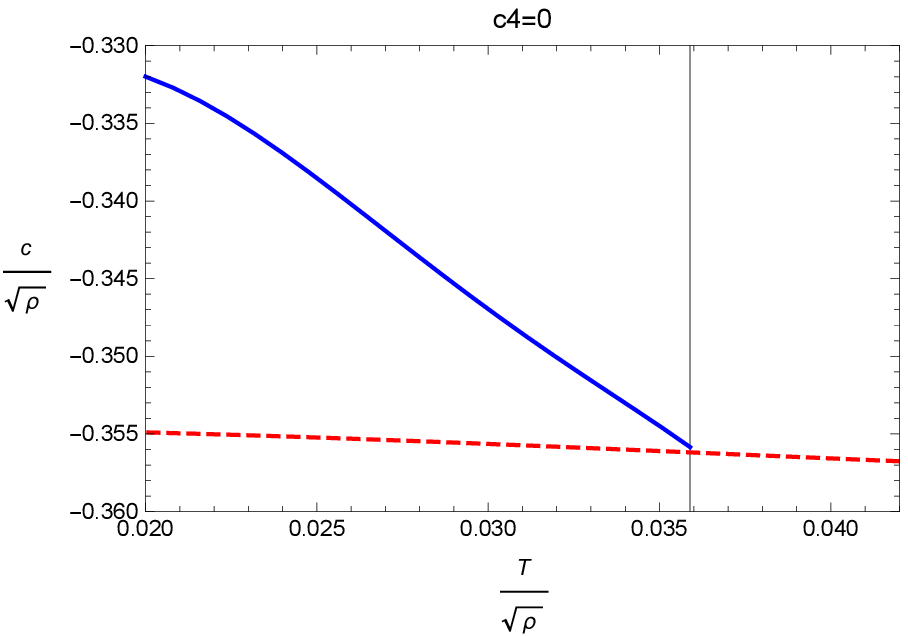}\hspace{1cm}
\includegraphics[scale=0.6]{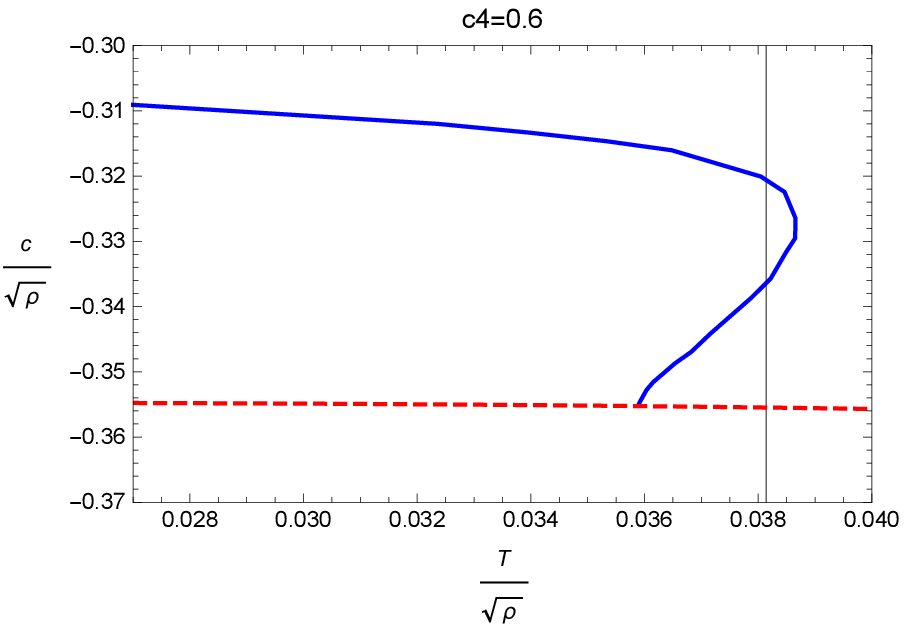}
\caption{\label{fighc} The holographic complexity  behaves as temperature for $c_4=0$ and $c_4=0.6$.}}
\end{figure}

We  turn to study the properties of  holographic complexity shown in figure \ref{fighc}. Different from the entanglement entropy, the complexity grows up as the temperature decreases which is similar as that found in \cite{Zangeneh:2017tub}.
For $c_4=0$, the holographic complexity in the second order phase transition case increases monotonously as  $T$ becomes lower, until $T$ reaches the critical value, then $c$ in superconducting state is always larger than that in normal state. Similarly, in the case of first order phase transition with $c_4=0.6$, When the temperature decreases, the numerical holographic complexity smoothly increases firstly and  presents multi-values  until the transformation point $T = 0.03559$. But the real process of  complexity has a jump to the maximal complexity at  $T_c=0.038143$ in the superconducting state.

Thus, we argue that the drop of the entanglement entropy and the jump of the  complexity at the critical point may be quite general features for the first
order superconducting  phase transition, while for second order phase transition, both of them are continue. Thus, both of holographic entanglement entropy and complexity  can be a possible probe of  phase transition in our superconductor model.
This numerical similarity is expected as we have mentioned in last subsection that  the features of entanglement entropy and
complexity are both determined by  the  features of extremal geometry.

\section{conclusion and discussion}

In this paper, we firstly reviewed  the holographic superconductor phase transition in the St$\ddot{u}$ckelberg  model with $\mathcal{F}(\psi)=\psi^{2}+c_{4}\psi^{4}$.   For the value of coupling parameter is less than the critical value, there is the second order phase transition; or else, when the coupling parameter is larger than the critical one, the first order phase transition  occurs. According to the analysis of free energy for the system, it is obvious that the system is experiencing more drastic condensation during the first order phase transition. Comparing to the critical temperature of phase transition, for the second order phase transition, the system reflects lower but has more stable critical temperature which is not affected  by the coupling strength.  For the first order case, the relation between critical temperature and coupling  strength is almost linearly increasing. Our results are similar to that found in the previous literatures\cite{Franco:2009if,Peng:2014ira}.

\begin{figure}[ht]
\center{
\includegraphics[scale=0.6]{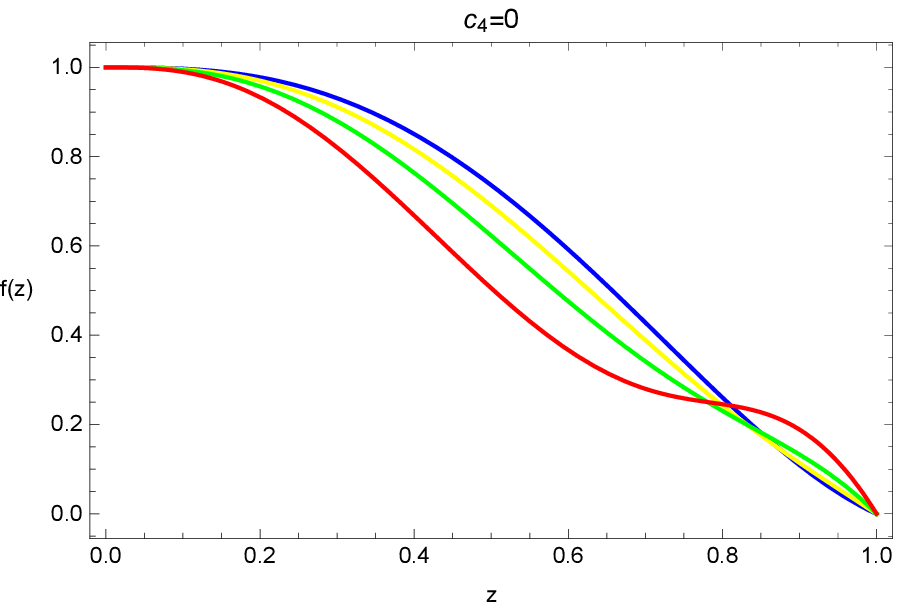}\hspace{1cm}
\includegraphics[scale=0.6]{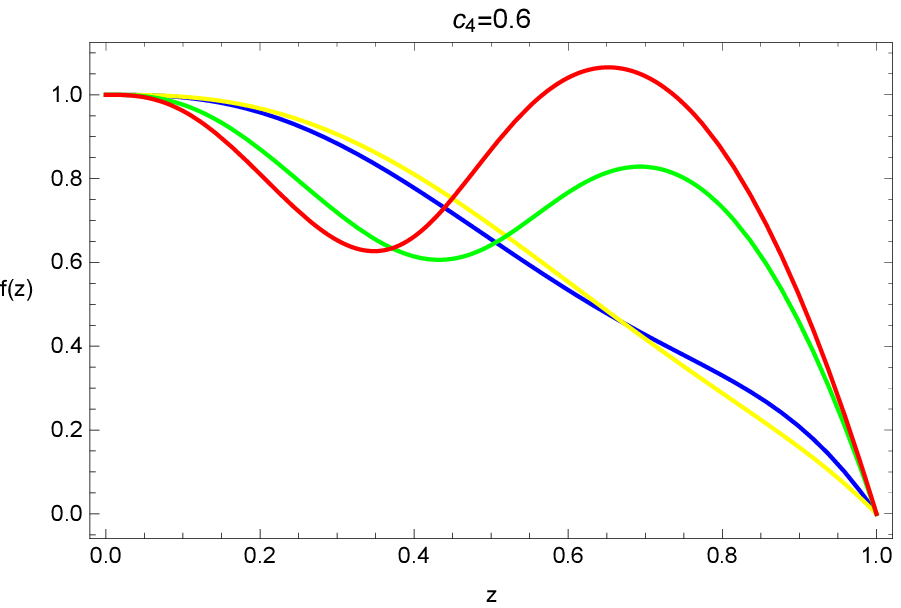}
\caption{\label{figsolution} Left: The profile of $f(z)$ with $c_4=0$. The different colors blue, yellow, green and red correspond to $T_c$,$0.8T_c$,$0.6T_c$ and $0.4T_c$, respectively;
Right: The profile of $f(z)$ with $c_4=0.6$.  The different colors  blue, yellow, green and red correspond to  $1.01T_c$, $T_c$, $0.80T_c$ and $0.70T_c$, respectively.}}
\end{figure}

Then  we studied the holographic entanglement entropy and holographic  complexity for the two types of phase transition. For  the second order phase transition, the holographic entanglement entropy  monotonously decreases while the complexity increases  as the temperature  becomes lower. While they have a multivalue area in high temperature region in first order case, which  appears in the same temperature region in free energy diagram. One of the multivalue lines is not  physical and does not exist in the real system.
According to \eqref{entanglement2} and \eqref{comple},  the behavior of the metric function $f(z)$ takes charge of the multivalue region of holographic entanglement entropy and complexity.
We show the profiles of $f(z)$ for samples of temperatures in both cases in figure \ref{figsolution}. It is obvious that in the first order case, the metric function $f(z)$ has unusual behaviors in low temperature like the case in \cite{Albash:2012pd}.

Both holographic entanglement entropy and complexity behaves differently crossing the second order and first order superconducting phase transition point. At the second order phase transition point, they are continuous and the slopes in terms of the temperature have a jump.  While in the first order superconducting phase transition, there exists  the drop for the entanglement entropy and the jump for the  complexity at the critical temperature. Thus, we argued that both of them can be used as  a probe to the order of  phase transition in our holographic superconducting model.
The discontinued behaviour of the slope around the transition temperature means a reorganization of the degrees of freedom in the system. This is a characteristic of second order phase transition since the entanglement entropy encodes the freedom degrees of the system. However, as we see that the situation is more complicated in first order phase transition. The existence of discontinuity in the behavior possibly indicates that not only the entangled pairs contributes to the entanglement entropy, but  extra  degrees of freedom  turning from other entanglement  are also reorganized. This extra pseudo-degree of freedom may come from the quantum effect, for instance, the thermodynamic system is excited to higher energy levels and certain symmetry is broken,  released in the strongly coupled system\cite{Albash:2010mv}. Thus, the new scales may  be detectable by holographic entanglement entropy and complexity, but the physical origins and deep insight from the field theory side deserve further study.
We note that the complexity is deeply connected with fidelity susceptibility,  which
is known to be able to probe phase transition\cite{Flory:2017ftd,Gan:2017qkz,MIyaji:2015mia,Moosa:2018mik}. It would be very interesting to pursuit the deep physics of difference in the probes by
studying  the fidelity susceptibility  in holographic superconducting models.

In this paper, we mainly focused on the difference of the qualitative  behavior of holographic complexity for different orders of superconducting phase transitions. Very recently, in the paper \cite{Yang:201902}, the authors studied the scaling of complexity on the temperature effected by the superconductor model parameters,  as well as the time dependent complexity when the system undergoes second order phase transition.  Study on these phenomena in superconducting models with different types of phase transition so as to further understand the deep physics in holographic superconductor  should be another interesting direction. We note that the time evolution of complexity under a quench in normal state have also been studied in \cite{Moosa:2017yvt,Chen:2018mcc,Chapman:2018dem,Ling:2018xpc,Fan:2018xwf,Ageev:2019fxn}.

It is noticed that here we computed the holographic complexity via complexity$=$volume conjecture and resolved its UV divergence tactfully.  As is known that another
 independent proposal to evaluate the complexity is  complexity$=$action conjecture \cite{Brown:2015bva,Brown:2015lvg,HosseiniMansoori:2018gdu}.  Then an important question involved is whether
 the physical features we found depends on the chosen conjecture. Especially, it was pointed out in \cite{Carmi:2016wjl,Moosa:2017yiz} that new UV divergences may arise in complexity$=$action conjecture.
 We hope to study the computation via complexity$=$action conjecture and do the comparison between the results from the two conjectures in the superconducting model in the near future.

The holographic entanglement and complexity are determined by the near horizon geometry. Thus, another possible direction is to analytical study their near horizon behavior to judge the order of phase. For example, one can study the holographic entanglement and complexity over the St$\ddot{u}$ckelberg superconductor at zero temperature limit (zero temperature limit of minimal coupling holographic superconductor can be seen in \cite{Horowitz:2009ij}) which may provide more insights to judge the order of superconduction phase transition.

\begin{acknowledgments}
We appreciate  Mahdis Ghodrati and Yan Peng for helpful correspondence. This work is supported by the Natural Science
Foundation of China under Grants No. 11705161 and No.11835009.
X. M. Kuang is also supported by Natural Science Foundation of Jiangsu Province under Grant No.BK20170481.
\end{acknowledgments}

\end{document}